\documentclass[pra,twocolumn,superscriptaddress,shownopacs,amssymb]{revtex4-2}
\usepackage{graphicx,psfrag,amsmath,amssymb}

\usepackage[usenames, dvipsnames]{color}

\usepackage{hyperref} 
\hypersetup{
    colorlinks=true,
    linkcolor=Blue,
    citecolor=Blue,
    filecolor=Blue,      
    urlcolor=Blue,
    pdftitle={Topological two-body bands},
    }

\begin{document}

\title{Topological two-body bands in a multiband Hubbard model}

\author{M. Iskin}
\affiliation{
Department of Physics, Ko\c{c} University, Rumelifeneri Yolu, 
34450 Sar\i yer, Istanbul, Turkey
}

\date{\today}

\begin{abstract}

In a multiband Hubbard model the self-consistency relations for the two-body 
bound-state bands are in the form of a nonlinear eigenvalue problem. Assuming 
that the resultant eigenvectors form an orthonormal set, e.g., in the strong-binding 
regime, here we reformulate their Berry curvatures and the associated Chern numbers. 
As an illustration we solve the two-body problem in a Haldane-Hubbard model with 
attractive onsite interactions, and analyze its topological phase diagrams from 
weak to strong couplings, i.e., by keeping track of the gap closings in between 
the low-lying two-body bands. The resultant Chern numbers are consistent with the 
lobe structure of the phase diagrams in the strong-coupling regime. 

\end{abstract}

\maketitle

\section{Introduction}
\label{sec:intro}

Topological classification of Bloch bands provides a fresh perspective on 
modern band theory, with important experimental implications~\cite{bansil16}.
For instance the Haldane model on a honeycomb lattice stabilizes quantum Hall 
effect by breaking both time-reversal and inversion symmetries through 
complex-valued hoppings and sublattice potential, i.e., without the need for the
Landau levels that are induced by an external magnetic field~\cite{haldane88}. 
The model features topologically distinct phases of matter with non-zero 
Chern numbers, making it one of the main workhorses for theoretical research 
on topological insulators and superconductors. Furthermore its experimental realization 
using an optical honeycomb lattice is a significant breakthrough in the field 
of topological matter as the atomic systems offer unprecedented control 
over the model parameters~\cite{jotzu14}.
Therefore a wide range of topological phases and their associated phenomena 
are within experimental reach, including the interplay between interactions 
and topology~\cite{cooper19}.

Exploring and discovering exotic phases of interacting matter such as fractional 
Chern and topological Mott insulators remains a primary objective in this 
field~\cite{liuz22, rachel18}. 
However, due to the complexity of interacting many-body problems, a bottom-up 
approach examining the exactly solvable two-body problem in a multiband Hubbard 
model can sometimes be useful~\cite{okuma23}. There are many recent works on
various topological aspects of the two-body 
problem~\cite{guo11, gorlach17, marques18, salerno18, lin20b, zurita20, salerno20, pelegri20}.
Among them the topological two-boson bound states in the repulsive Haldane-Bose-Hubbard 
model were analyzed using exact diagonalization in real space~\cite{salerno18}, 
and in this paper, we examine the two-body problem for a multiband Fermi-Hubbard 
model in momentum space. The self-consistency relations for the two-body 
bound-state bands are in the form of a nonlinear eigenvalue 
problem~\cite{iskin21, iskin22f, iskin22d}, and we reformulate the Berry curvature 
and the associated Chern number in the strong-binding regime with the underlying 
assumption that the resultant eigenvectors form approximately an orthonormal set. 
As an illustration we construct topological phase diagrams for the attractive 
Haldane-Hubbard model from weak to strong couplings, and show that the lobe 
structure of the phase diagram is consistent with the associated Chern numbers 
in the strong-coupling regime. 
We would like to emphasize that our formulation is also pertinent to various 
other physical systems, e.g., in the investigation of flat-band physics in 
Kagome metals and twisted bilayer graphene, where recent advances in the study 
of strong-correlation physics have already revealed intriguing connections 
between topology and electronic properties~\cite{yin22, kang20, balents20, herzog22}.

The rest of the paper is organized as follows. In Sec.~\ref{sec:tbbs} we express 
the exact solution in the form of a nonlinear eigenvalue problem, and perform its 
strong-coupling expansion. In Sec.~\ref{sec:em} we derive an effective Hamiltonian 
for the strongly-bound pairs in a generic two-band model, and benchmark it  with 
the Haldane, Su-Schrieffer-Heeger and Hofstadter models. In Sec.~\ref{sec:tpd} we 
focus on the Haldane-Hubbard model, and analyze its topological phase diagram numerically 
for the two-body bands. The paper ends with a brief summary of our conclusions in 
Sec.~\ref{sec:conc}, and three Appendices on (A) derivation of the hopping parameters
for the effective Hamiltonian, (B) derivation of the Berry curvature for the 
eigenvectors of the nonlinear eigenvalue problem and (C) application to the 
isolated flat bands.

\section{Two-body bound states}
\label{sec:tbbs}

In this section we consider a generic tight-binding lattice with multiple 
sublattices, and show that the number of sublattices determines not only 
the number of Bloch bands but also the number of so-called two-body bound-state 
bands as follows.

\subsection{One-body problem}
\label{sec:obp}

The one-body problem in a multiband lattice is described by
\begin{align}
h_{\sigma \mathbf{k}} \mathbf{f}_{n \sigma \mathbf{k}} 
= \varepsilon_{n \sigma \mathbf{k}} \mathbf{f}_{n \sigma \mathbf{k}},
\end{align}
where the matrix $h_{\sigma \mathbf{k}}$ represents the Bloch Hamiltonian for the 
spin-$\sigma$ particle with momentum $\mathbf{k}$ in the sublattice basis 
$\alpha = \{A, B, C, \cdots \}$, its eigenvectors
$
\mathbf{f}_{n \sigma \mathbf{k}} = (n_{A \sigma \mathbf{k}}, n_{B \sigma \mathbf{k}}, 
n_{C \sigma \mathbf{k}}, \cdots)^\mathrm{T}
$
with
$
n_{\alpha \sigma \mathbf{k}} = \langle \alpha |n \sigma \mathbf{k} \rangle
$
represent the periodic part of the Bloch states $|n \sigma \mathbf{k} \rangle$ 
and its eigenvalues $\varepsilon_{n \sigma \mathbf{k}}$ determine the Bloch bands.
Here $\mathrm{T}$ is the transpose which is in such a way that
$
\sum_{\beta} h^{\alpha \beta}_{\sigma \mathbf{k}} n_{\beta \sigma \mathbf{k}} 
= \varepsilon_{n \sigma \mathbf{k}} n_{\alpha \sigma \mathbf{k}}.
$
In the presence of two sublattices only, the Bloch Hamiltonian can be written as 
\begin{align}
h_{\sigma \mathbf{k}} &= d_{\sigma \mathbf{k}}^0 \tau_0 
+ \mathbf{d}_{\sigma \mathbf{k}} \cdot \boldsymbol{\tau},
\end{align}
where the $\mathbf{k}$ dependences of $d_{\sigma \mathbf{k}}^0$ and
$
\mathbf{d}_{\sigma \mathbf{k}} 
= (d_{\sigma \mathbf{k}}^x, d_{\sigma \mathbf{k}}^y, d_{\sigma \mathbf{k}}^z)
$
are determined by the details of the hopping processes and onsite energies. 
Here $\tau_0$ is an identity matrix and 
$
\boldsymbol{\tau} = (\tau_x, \tau_y, \tau_z)
$ 
is a vector of Pauli spin matrices for the sublattice sector. The Bloch bands are given by
$
\varepsilon_{s \sigma \mathbf{k}} = d_{\sigma \mathbf{k}}^0 + s d_{\sigma \mathbf{k}},
$
where $s = \pm$ denotes the upper and lower bands, and
$
d_{\sigma \mathbf{k}} = \sqrt{(d_{\sigma \mathbf{k}}^x)^2 + (d_{\sigma \mathbf{k}}^y)^2 
+(d_{\sigma \mathbf{k}}^z)^2}
$
is the magnitude of $\mathbf{d}_{\sigma \mathbf{k}}$. The sublattice projections 
of Bloch states 
$
\mathbf{f}_{s \sigma \mathbf{k}} = (s_{A \sigma \mathbf{k}}, s_{B \sigma \mathbf{k}})^\mathrm{T}
$
can be written as
$
\mathbf{f}_{+, \sigma \mathbf{k}} = \big( u_{\sigma \mathbf{k}}, 
v_{\sigma \mathbf{k}} e^{i\varphi_{\sigma \mathbf{k}}} \big)^\mathrm{T}
$
for the upper band and
$
\mathbf{f}_{-, \sigma \mathbf{k}} = \big( -v_{\sigma \mathbf{k}} e^{-i\varphi_{\sigma \mathbf{k}}}, 
u_{\sigma \mathbf{k}} \big)^\mathrm{T}
$
for the lower band, where 
$
u_{\sigma \mathbf{k}} = \sqrt{(d_{\sigma \mathbf{k}} 
+ d_{\sigma \mathbf{k}}^z)/(2d_{\sigma \mathbf{k}})}
$
and
$
v_{\sigma \mathbf{k}} = \sqrt{(d_{\sigma \mathbf{k}} 
- d_{\sigma \mathbf{k}}^z)/(2d_{\sigma \mathbf{k}})}
$
are the usual amplitudes and
$
\varphi_{\sigma \mathbf{k}} = \arg(d_{\sigma \mathbf{k}}^x + id_{\sigma \mathbf{k}}^y)
$
is the polar angle on the Bloch sphere.
Using the solutions of the one-body problem, next we construct solutions for 
the low-lying two-body bound states.

\subsection{Nonlinear eigenvalue problem}
\label{sec:nep}

The two-body problem in a multiband Hubbard model is exactly solvable, and 
the resultant spectrum can be divided into three distinct set of 
solutions~\cite{iskin21, iskin22f, iskin22d}.
The first set is the scattering continua, and these states correspond to 
two unbound (non-interacting) particles. There are $M_b(M_b+1)/2$ possible continua 
in total where $M_b$ is the number of Bloch bands, i.e., the number of sublattices.
The second set is the so-called offsite bound states, and they lie in between 
the scattering continua. For this reason these states remain weakly bound even 
in the strongly-interacting regime.
The third set is the so-called onsite bound states, and they lie either on top 
or at the bottom of the two-body spectrum depending on whether the onsite Hubbard 
interaction is repulsive or attractive, respectively. There are $M_b$ of them for 
a given center of mass momentum $\mathbf{q}$ of the two particles.
These states become strongly bound in the strongly-interacting regime, where 
they eventually correspond to strongly-localized onsite pairs. In this paper we 
focus only on this last set of solutions because they give rise to the 
two-body bands as a function of $\mathbf{q}$.

For the two-body problem between an $\uparrow$ and a $\downarrow$ fermion, 
the third set can be determined entirely via the self-consistency 
relation~\cite{iskin21, iskin22f}
\begin{align}
\label{eqn:Nalpha}
N_{\alpha \mathbf{q}} = \frac{U}{M_c} \sum_{nm\mathbf{k} \beta}
\frac{
n_{\beta \uparrow \mathbf{k+q}}^* m_{\beta \downarrow -\mathbf{k}}^*
m_{\alpha \downarrow -\mathbf{k}} n_{\alpha \uparrow \mathbf{k+q}}}
{\varepsilon_{n \uparrow \mathbf{k}+\mathbf{q}} 
+ \varepsilon_{m \downarrow -\mathbf{k}} - E_{N \mathbf{q}}}
N_{\beta \mathbf{q}},
\end{align}
where $U \ge 0$ is the strength of the attractive onsite Hubbard interaction,
$M_c$ is the number of unit cells in the lattice, and $E_{N \mathbf{q}}$ is 
the energy of the bound state.
This expression is also valid for $U < 0$, in which case $-U$ corresponds to the 
strength of the repulsive onsite Hubbard interaction. One can rewrite it as 
$
\mathbf{G}_{N \mathbf{q}} \mathbf{F}_{N \mathbf{q}} = 0,
$
and determine $E_{N \mathbf{q}}$ self-consistently through an iterative approach,
where 
$
\mathbf{F}_{N \mathbf{q}} =  \big( N_{A \mathbf{q}}, N_{B \mathbf{q}},
N_{C \mathbf{q}}, \cdots \big)^\mathrm{T}
$
represents the bound state $|N \mathbf{q} \rangle$ in the sublattice basis.
For a given $\mathbf{q}$, different $N$ corresponds to a self-consistent solution 
that is determined by setting the first, or second, or third, etc., eigenvalue 
of the Hermitian matrix $\mathbf{G}_{N \mathbf{q}}$ to be 0. Note that each 
$E_{N \mathbf{q}}$ solution gives in return a different $\mathbf{G}_{N \mathbf{q}}$ 
matrix once the self-consistency is achieved.

Equation~(\ref{eqn:Nalpha}) can also be interpreted as a nonlinear eigenvalue problem
\begin{align}
\label{eqn:HNq}
H_{N \mathbf{q}} \mathbf{F}_{N \mathbf{q}} 
= E_{N \mathbf{q}} \mathbf{F}_{N \mathbf{q}},
\end{align}
in such a way that
$
\sum_{\beta} H^{\alpha \beta}_{N \mathbf{q}} N_{\beta \mathbf{q}} 
= E_{N \mathbf{q}} N_{\alpha \mathbf{q}}.
$
This eigenvalue problem is not in the usual form because the matrix elements 
\begin{align}
\label{eqn:Hmatrix}
H^{\alpha \beta}_{N \mathbf{q}} = \frac{E_{N \mathbf{q}} U} {M_c} \sum_{nm\mathbf{k}}
\frac{n_{\beta \uparrow \mathbf{k+q}}^* m_{\beta \downarrow -\mathbf{k}}^*
m_{\alpha \downarrow -\mathbf{k}} n_{\alpha \uparrow \mathbf{k+q}}}
{\varepsilon_{n \uparrow \mathbf{k}+\mathbf{q}} 
+ \varepsilon_{m \downarrow -\mathbf{k}} - E_{N \mathbf{q}}}
\end{align}
depend explicitly on the eigenvalue $E_{N \mathbf{q}}$, and hence, it corresponds 
to a self-consistency relation for each $E_{N \mathbf{q}}$.
For instance, in the presence of two sublattices only, the relevant matrix for 
a given bound-state solution $E_{S \mathbf{q}}$ can be written as
\begin{align}
\label{eqn:HSq}
H_{S \mathbf{q}} = D_{S \mathbf{q}}^0 \tau_0 
+ \mathbf{D}_{S \mathbf{q}} \cdot \boldsymbol{\tau},
\end{align}
where
$
D_{S \mathbf{q}}^0 = (H^{AA}_{S \mathbf{q}} + H^{BB}_{S \mathbf{q}})/2,
$
$
D_{S \mathbf{q}}^x = \mathrm{Re} H^{BA}_{S \mathbf{q}},
$
$
D_{S \mathbf{q}}^y = \mathrm{Im} H^{BA}_{S \mathbf{q}},
$
and 
$
D_{S \mathbf{q}}^z = (H^{AA}_{S \mathbf{q}} - H^{BB}_{S \mathbf{q}})/2.
$
Here $\mathrm{Re}$ and $\mathrm{Im}$ denotes, respectively, the real and imaginary parts.
The corresponding eigenvector of $H_{S \mathbf{q}}$ that satisfies the self-consistency
relation can be denoted as
$
\mathbf{F}_{S \mathbf{q}} =  \big( S_{A \mathbf{q}}, S_{B \mathbf{q}} \big)^\mathrm{T}.
$
Note that, for a given self-consistent solution $E_{S \mathbf{q}}$, the matrix 
$H_{S \mathbf{q}}$ has two eigenvalues, but only one of those satisfies the 
self-consistency relation. The other solution and its eigenvector are irrelevant.
Alternatively the self-consistency relations can be written as
$
E_{S \mathbf{q}} = D_{S \mathbf{q}}^0 + S' D_{S \mathbf{q}},
$
where $S' = \pm$ and
$
D_{S \mathbf{q}} = \sqrt{(D_{S \mathbf{q}}^x)^2 + (D_{S \mathbf{q}}^y)^2 
+(D_{S \mathbf{q}}^z)^2}.
$

For a given $\mathbf{q}$, since each $E_{N \mathbf{q}}$ solution is associated with 
a different Hermitian matrix $H_{N \mathbf{q}}$, the corresponding eigenvectors 
$\mathbf{F}_{N \mathbf{q}}$ do not necessarily form an orthonormal set in general.
The only exception for this seems to be the strong-binding regime, where the onsite 
bound states become strongly localized on a single lattice site, i.e., 
on one of the sublattices, and become approximately orthogonal to each other 
at finite $|U|$. 
Note that there are as many two-body bands as the number of sublattices or 
equivalently as the number of Bloch bands. Thus it may be possible to interpret 
$H_{N \mathbf{q}}$ as an effective Hamiltonian for the onsite bound states as 
discussed next.

\subsection{Strong-binding regime}
\label{sec:sbr}

As an illustration here we focus on lattices with a two-point basis for the simplicity 
of their presentation. A similar analysis can be performed for multiband lattices.
It turns out the binding energy is of order $|U|$ in the strong-binding 
regime when $|U|$ is much larger than the bandwidth of the lowest Bloch band. 
In general strong binding requires strong interactions in the case of dispersive 
Bloch bands. However, in the particular case when the lowest (highest) Bloch 
band is flat and it is separated from the other bands by an energy gap, 
even an arbitrarily small $U > 0$ ($U < 0$) can be treated as strong-binding regime. 
In such a case the binding energy of the lowest bound state is known to be of 
order $|U|/M_b$. Thus isolated flat bands are also amenable to a similar 
strong-binding expansion when the interactions are weak.

In the strong-coupling regime, the matrix elements of Eq.~(\ref{eqn:HSq}) can 
be expanded as
\begin{align}
D_{S \mathbf{q}}^0 &= -U \bigg(1 + \frac{\lambda_{1 \mathbf{q}}}{E_{S \mathbf{q}}} 
+ \frac{\lambda_{2 \mathbf{q}}}{E_{S \mathbf{q}}^2}
+ \frac{\lambda_{3 \mathbf{q}}}{E_{S \mathbf{q}}^3} + \cdots \bigg), \\
D_{S \mathbf{q}}^x &+ i D_{S \mathbf{q}}^y = - U \bigg( \frac{\kappa_{2 \mathbf{q}}}{E_{S \mathbf{q}}^2} 
+ \frac{\kappa_{3 \mathbf{q}}}{E_{S \mathbf{q}}^3} + \cdots \bigg), \\
D_{S \mathbf{q}}^z &= - U \bigg( \frac{\gamma_{1 \mathbf{q}}}{E_{S \mathbf{q}}} 
+ \frac{\gamma_{2 \mathbf{q}}}{E_{S \mathbf{q}}^2} 
 + \frac{\gamma_{3 \mathbf{q}}}{E_{S \mathbf{q}}^3} + \cdots \bigg),
\end{align}
where $\lambda_{i \mathbf{q}}$ and $\gamma_{i \mathbf{q}}$ are real numbers but 
$\kappa_{i \mathbf{q}}$ is a complex number. After some algebra discussed in Appendix A, 
these expansion coefficients can be written as
\begin{align}
\lambda_{1 \mathbf{q}} &= \frac{1}{M_c} \sum_\mathbf{k} \big( 
d_{\uparrow \mathbf{k+q}}^0 + d_{\downarrow -\mathbf{k}}^0 \big), \\
\lambda_{2 \mathbf{q}} &= \frac{1}{M_c} \sum_\mathbf{k} \big[ 
\big( d_{\uparrow \mathbf{k+q}}^0 + d_{\downarrow -\mathbf{k}}^0 \big)^2 
+ \big( d_{\uparrow \mathbf{k+q}}^z + d_{\downarrow -\mathbf{k}}^z \big)^2 \nonumber \\
& \;\;\;\;\;\;\;\;\;\;\;\;\;\;\;\;\; + |g_{\uparrow \mathbf{k+q}}|^2 + |g_{\downarrow -\mathbf{k}}|^2 \big], \\
\kappa_{2 \mathbf{q}} &= \frac{2}{M_c} \sum_\mathbf{k} g_{\uparrow \mathbf{k+q}} g_{\downarrow -\mathbf{k}}, \\
\gamma_{1 \mathbf{q}} &= \frac{1}{M_c} \sum_\mathbf{k} 
\big( d_{\uparrow \mathbf{k+q}}^z + d_{\downarrow -\mathbf{k}}^z \big), \\
\gamma_{2 \mathbf{q}} &= \frac{2}{M_c} \sum_\mathbf{k} 
\big( d_{\uparrow \mathbf{k+q}}^0 + d_{\downarrow -\mathbf{k}}^0 \big) 
\big( d_{\uparrow \mathbf{k+q}}^z + d_{\downarrow -\mathbf{k}}^z \big),
\end{align}
where
$
g_{\sigma \mathbf{k}} = d_{\sigma \mathbf{k}}^x + i d_{\sigma \mathbf{k}}^y
= |g_{\sigma \mathbf{k}}|e^{i \varphi_{\sigma \mathbf{k}}}
$
is defined for convenience. Typically $\lambda_{1 \mathbf{q}} = 0$ when the 
sublattice potentials are symmetric around $0$, and
$
\gamma_{1 \mathbf{q}} = 0 = \gamma_{2 \mathbf{q}}
$
when the Bloch Hamiltonian exhibits time-reversal symmetry.
See Sec.~\ref{sec:em} for example models.

When $\gamma_{1 \mathbf{q}} \ge 0$, the bound-state energies are determined by 
the self-consistency relation
$
E_{S \mathbf{q}} = D_{S \mathbf{q}}^0 + S D_{S \mathbf{q}},
$
where $S = \pm$ corresponds to upper and lower two-body bands, respectively. On the other hand
$S = \pm$ corresponds to lower and upper two-body bands when $\gamma_{1 \mathbf{q}} < 0$.
Up to first order in $1/U$, the self-consistency relations lead to
$
E_{S \mathbf{q}} = -U + \epsilon_{S \mathbf{q}} + \eta_{S \mathbf{q}}/U + \mathcal{O}(1/U^2),
$
where $\epsilon_{S \mathbf{q}} = \lambda_{1 \mathbf{q}} + S|\gamma_{1 \mathbf{q}}|$, 
and
$
\eta_{S \mathbf{q}} = (\lambda_{1 \mathbf{q}}^2 - \lambda_{2 \mathbf{q}} 
+ S \sqrt{\gamma_{2 \mathbf{q}}^2+|\kappa_{2 \mathbf{q}}|^2})/U
$
when $\gamma_{1 \mathbf{q}} = 0$ but
$
\eta_{S \mathbf{q}} = \epsilon_{S \mathbf{q}}^2 - \lambda_{2 \mathbf{q}} 
- S \gamma_{2 \mathbf{q}} |\gamma_{1 \mathbf{q}}|/\gamma_{1 \mathbf{q}}
$
when $\gamma_{1 \mathbf{q}} \ne 0$.
Up to second order in $1/U$, thus we obtain
%%%
\begin{align}
\label{eqn:D0U}
D_{S \mathbf{q}}^0 &= -U + \lambda_{1 \mathbf{q}} 
+ \frac{\lambda_{1 \mathbf{q}} \epsilon_{S \mathbf{q}} - \lambda_{2 \mathbf{q}}}{U} \nonumber \\ 
&+ \frac{\lambda_{1 \mathbf{q}}(\eta_{S \mathbf{q}} + \epsilon_{S \mathbf{q}}^2) 
- 2\lambda_{2 \mathbf{q}} \epsilon_{S \mathbf{q}} + \lambda_{3 \mathbf{q}}}{U^2}  + \cdots, \\
\label{eqn:DxyU}
D_{S \mathbf{q}}^x &+ iD_{S \mathbf{q}}^y = -\frac{\kappa_{2 \mathbf{q}}}{U} 
- \frac{2\kappa_{2 \mathbf{q}} \epsilon_{S \mathbf{q}} - \kappa_{3 \mathbf{q}}}{U^2} + \cdots, \\
\label{eqn:DzU}
D_{S\mathbf{q}}^z &= \gamma_{1 \mathbf{q}} 
+ \frac{\gamma_{1 \mathbf{q}} \epsilon_{S \mathbf{q}} - \gamma_{2 \mathbf{q}}}{U} \nonumber \\ 
&+ \frac{\gamma_{1 \mathbf{q}}(\eta_{S \mathbf{q}} + \epsilon_{S \mathbf{q}}^2) 
- 2\gamma_{2 \mathbf{q}} \epsilon_{S \mathbf{q}} + \gamma_{3 \mathbf{q}}}{U^2} + \cdots.
\end{align}
These expressions can be used to construct an effective 
Hamiltonian $H_\mathbf{q}$ for the onsite bound states in the strong-coupling regime. 
For instance, when Eqs.~(\ref{eqn:D0U}), (\ref{eqn:DxyU}) and (\ref{eqn:DzU}) do not 
depend on $S$ (e.g. when $\lambda_{1 \mathbf{q}} = 0 = \gamma_{1 \mathbf{q}}$), 
$H_\mathbf{q}$ coincides trivially with $H_{+, \mathbf{q}} = H_{-, \mathbf{q}}$ 
up to second order in $1/U$.
As a non-trivial illustration, next we derive $H_\mathbf{q}$ up to first order in 
$1/U$ for a generic lattice using perturbation theory.

\section{Example models}
\label{sec:em}

Suppose $\gamma_{1 \mathbf{q}} \ge 0$ without losing generality so that $S = \pm$ 
corresponds to upper and lower two-body bands, respectively, and the associated 
onsite bound states are strongly localized on sublattice A and B, respectively. 
This is clearly seen in the unperturbed (i.e., $|U| \to \infty$) problem, where 
sublattices A and B are decoupled from each other [i.e., Eq.~(\ref{eqn:DxyU}) $\to$ 0)], 
and the unperturbed two-body band
$
E_{S \mathbf{q}}^{(0)} = -U + \lambda_{1 \mathbf{q}} + S \gamma_{1 \mathbf{q}}
$
corresponds to a completely localized state on the relevant sublattice. 
Then the finite-$U$ effects can be taken into account through perturbation theory.
For instance, at first order in $1/U$, the matrix elements $H_\mathbf{q}^{\alpha\beta}$ 
of the effective Hamiltonian $H_\mathbf{q}$ are such that
$
H_\mathbf{q}^{AA} \equiv H_{+, \mathbf{q}}^{AA} = D_{+, \mathbf{q}}^0 + D_{+, \mathbf{q}}^z,
$ 
$
H_\mathbf{q}^{BB} \equiv H_{-, \mathbf{q}}^{BB} = D_{-, \mathbf{q}}^0 - D_{-, \mathbf{q}}^z,
$ 
and
$
H_\mathbf{q}^{BA} \equiv H_{\pm, \mathbf{q}}^{BA}  = D_{\pm, \mathbf{q}}^x + iD_{\pm, \mathbf{q}}^y.
$ 
This effective Hamiltonian can be written as
$
H_\mathbf{q} = D_\mathbf{q}^0 \tau_0 + \mathbf{D}_\mathbf{q} \cdot \boldsymbol{\tau},
$
where
\begin{align}
D_\mathbf{q}^0 &= -U + \lambda_{1 \mathbf{q}} 
+ \frac{\lambda_{1 \mathbf{q}}^2 + \gamma_{1 \mathbf{q}}^2 - \lambda_{2 \mathbf{q}}}{U}, \\
D_\mathbf{q}^x &+ iD_\mathbf{q}^y = -\frac{\kappa_{2 \mathbf{q}}}{U}, \\
D_\mathbf{q}^z &= \gamma_{1 \mathbf{q}} + \frac{2\lambda_{1 \mathbf{q}} \gamma_{1 \mathbf{q}} - \gamma_{2 \mathbf{q}}}{U}
\end{align}
determine its matrix elements. These expressions are readily applicable to any Bloch 
Hamiltonian with a two-point basis. Some important models are discussed next.

\subsection{Haldane model}
\label{sec:ham}

In the original Haldane model on a honeycomb lattice with a two-point basis, while the 
nearest-neighbor (i.e., inter-sublattice) hopping $t_{nn} = t$ is a real parameter, the 
next-nearest-neighbor (i.e., intra-sublattice) hopping  $t_{nnn} = t' e^{i\phi}$ 
is a complex parameter~\cite{haldane88}. Its Bloch Hamiltonian
$
h_{\sigma \mathbf{k}} \equiv h_{\mathbf{k}}
$
is such that 
$
d_\mathbf{k}^0 = -2t' \cos \phi \sum_{j = 1}^3 \cos (\mathbf{k} \cdot \boldsymbol{\nu_j}),
$
$
d_\mathbf{k}^x + id_\mathbf{k}^y = -t \sum_{j = 1}^3 e^{i \mathbf{k} \cdot \boldsymbol{e_j}}
$
and
$
d_\mathbf{k}^z = \delta -2t' \sin \phi \sum_{j = 1}^3 \sin (\mathbf{k} \cdot \boldsymbol{\nu_j}),
$
where $\delta$ is the onsite energy difference between sublattices. Here we define
$
\boldsymbol{e_1} = (0, a),
$
$
\boldsymbol{e_2} = (-\sqrt{3}a/2, -a/2) 
$
and
$
\boldsymbol{e_3} = - (\boldsymbol{e_1}+\boldsymbol{e_2}) =  (\sqrt{3}a/2, -a/2)
$
for the nearest-neighbor hoppings, and similarly
$
\boldsymbol{\nu_1} = (\sqrt{3}a, 0),
$
$
\boldsymbol{\nu_2} = (-\sqrt{3}a/2, 3a/2) 
$
and
$
\boldsymbol{\nu_3} = - (\boldsymbol{\nu_1}+\boldsymbol{\nu_2}) =  (-\sqrt{3}a/2, -3a/2)
$
for the next-nearest-neighbor hoppings, where $a$ is the lattice spacing.
Its Brillouin zone has the shape of a hexagon, and it is such that the K and 
K$^\prime$ valleys are at
$
\mathbf{K} = [4\pi/(3\sqrt{3}a), 0]
$
and
$
\mathbf{K'} = [2\pi/(3\sqrt{3}a), 2\pi/(3a)]
$
points, respectively.

After some tedious bookkeeping, one can show that the expansion coefficients for the 
two-body problem are
$
\lambda_{1 \mathbf{q}} = 0,
$
$
\lambda_{2 \mathbf{q}} = 6t^2 + 12t'^2 + 4\delta^2 + 4t'^2 \cos(2\phi) 
\sum_{j = 1}^3 \cos (\mathbf{q} \cdot \boldsymbol{\nu_j}),
$
$
\kappa_{2 \mathbf{q}} = 2t^2 \sum_{j = 1}^3 e^{i \mathbf{q} \cdot \boldsymbol{e_j}},
$
$
\gamma_{1 \mathbf{q}} = 2\delta
$
and
$
\gamma_{2 \mathbf{q}} = 4t'^2 \sin(2\phi) \sum_{j = 1}^3 \sin (\mathbf{q} \cdot \boldsymbol{\nu_j}).
$
Thus the effective Hamiltonian for the onsite bound states is described by
$
D_\mathbf{q}^0 = - U - \Lambda - 2T' \cos \Phi \sum_{j = 1}^3 \cos (\mathbf{q} \cdot \boldsymbol{\nu_j}),
$
$
D_\mathbf{q}^x + iD_\mathbf{q}^y = -T \sum_{j = 1}^3 e^{i \mathbf{q} \cdot \boldsymbol{e_j}}
$
and
$
D_\mathbf{q}^z = \Delta - 2T' \sin \Phi \sum_{j = 1}^3 \sin (\mathbf{q} \cdot \boldsymbol{\nu_j}),
$
where $\Lambda = (6t^2 + 12t'^2)/U$ is an onsite energy offset, $T = 2t^2/U$ is the effective 
nearest-neighbor hopping parameter, $T' = 2t'^2/U$ is the amplitude and $\Phi = 2\phi$ is the 
phase of the effective next-nearest-neighbor hopping parameter, and $\Delta = 2\delta$ is the 
effective onsite energy difference between sublattices. These effective parameters are consistent 
with the recent literature~\cite{salerno18}~\footnote{
As far as the two-body problem is concerned, there is no essential difference between the 
results of Bose-Hubbard and Fermi-Hubbard models when 
$
h_{\sigma \mathbf{k}} \equiv h_{\mathbf{k}}
$
in the latter~\cite{iskin22d, iskin22f}.
}.

Note that $\Lambda = 2 (z_{nn} t^2 + z_{nnn} t'^2)/U$, where 
$z_{nn} = 3$ is the nearest-neighbor coordination number and $z_{nnn} = 4$ 
is the next-nearest-neighbor coordination number.
The origin of such a $\mathbf{q}$-independent onsite energy offset is as follows. 
When a bound state breaks up at a cost of binding energy $U$ in the denominator, 
one of its constituents can hop to a neighboring site and then come back to the original 
site to recombine, leading to $t_\sigma^2 = t^2$ in the numerator. Thus the center of 
mass of the pair does not play a role in this process.
The coordination numbers appear because such a process can happen with all neighboring 
sites. The factor of 2 accounts for the spin.
On the other hand the effective hopping parameters are $\mathbf{q}$ dependent because 
when a bound state breaks up and one of its constituents hops to a neighboring site, 
the other particle follows it and hops to the same site, leading to 
$t_\uparrow t_\downarrow = t^2$ in the numerator. This is the only physical
mechanism for a strongly-bound pair of particles to move in the Hubbard model.

\subsection{Su-Schrieffer-Heeger-Hubbard model}
\label{sec:sshm}

In the Su–Schrieffer–Heeger model on a linear chain with a two-point basis, 
while the onsite energy difference between sublattices and the intra-sublattice 
hopping parameters are set to 0, the inter-sublattice hopping parameters alternate 
between $t_L$ and $t_R$ in the lattice~\cite{su79}. Its Bloch Hamiltonian
$
h_{\sigma \mathbf{k}} \equiv h_{\mathbf{k}}
$
is such that 
$
d_k^0 = 0 = d_k^z,
$
$
d_k^x = -t_L - t_R \cos(k a)
$
and
$
d_k^y = -t_R \sin(k a),
$
where $a$ is the lattice spacing, $-\pi/a \le k < \pi/a$ defines the Brillouin zone, 
and $t_L$ and $t_R$ are real hopping parameters to the left and right of sublattice A, 
respectively.
The Bloch bands can be written as
$
\varepsilon_{s k} = s\sqrt{t_L^2 + t_R^2 + 2t_L t_R \cos(k a)}.
$

After some simple algebra, one can show that the expansion coefficients for 
the two-body problem are
$
\lambda_{1 q} = 0 =\gamma_{1 q} = \gamma_{2 q}
$ 
due to particle-hole and time-reversal symmetries,
$
\lambda_{2 q} = 2t_L^2 + 2t_R^2
$
and
$
\kappa_{2 q} = 2t_L^2 + 2t_R^2 e^{i q a}.
$
Thus the effective Hamiltonian for the onsite bound states is described by
$
D_q^0 = - U - \Lambda,
$
$
D_q^x + iD_q^y = -T_L - T_R e^{iq a}
$
and
$
D_q^z = 0,
$
where $\Lambda = (2t_L^2 + 2t_R^2)/U$ is an onsite energy offset, and $T_L = 2t_L^2/U$ 
and $T_R = 2t_R^2/U$ are the effective nearest-neighbor hopping parameters.
These effective parameters are consistent with the recent literature~\cite{liberto16, lin20}.
Note that $\Lambda = 2 (z_L t_L^2 + z_R t_R^2)/U$, where 
$z_L = 1$ and $z_R = 1$ are the corresponding coordination numbers
to the left and to the right, respectively. Thus, similar to the Bloch bands, 
the two-body bands can be written as
$
E_{S q} = - U - \Lambda + S \sqrt{T_L^2 + T_R^2 + 2T_L T_R \cos(q a)}.
$

\subsection{Hofstadter-Hubbard model at $\alpha_{\sigma B} = 1/2$}
\label{sec:hom}

The Hofstadter model on a square lattice with nearest-neighbor hoppings $t$ is 
described by a two-point basis when the magnetic flux quanta per unit cell is 
$
\alpha_{\sigma B} = B_0 a^2/\phi_0 = 1/2
$ 
~\cite{hofstadter76, cocks12, shaffer21}.
Here $B_0$ is the strength of the perpendicular magnetic field, $a$ is the lattice 
spacing, and $\phi_0$ is the magnetix-flux quantum.
Its Bloch Hamiltonian
$
h_{\sigma \mathbf{k}} \equiv h_{\mathbf{k}}
$
is such that 
$
d_\mathbf{k}^0 = 0,
$
$
d_\mathbf{k}^x = -t - t\cos(2k_x a),
$
$
d_\mathbf{k}^y = - t\sin(2k_x a),
$
and
$
d_\mathbf{k}^z = 2t \cos(k_y a),
$
where $-\pi/(2a) \le k_x < \pi/(2a)$ and $-\pi/a \le k_y < \pi/a$ defines the
magnetic Brillouin zone. 

After some simple algebra, one can show that the expansion coefficients for 
the two-body problem are
$\lambda_{1 \mathbf{q}} = 0 = \gamma_{1 \mathbf{q}} = \gamma_{2 \mathbf{q}}$ 
due to particle-hole and time-reversal symmetries,
$
\lambda_{2 \mathbf{q}} = 8t^2 + 4t^2\cos(q_y a)
$
and
$
\kappa_{2 \mathbf{q}} = 2t^2 + 2t^2 e^{i2q_x a}.
$
Thus the effective Hamiltonian for the onsite bound states is described by
$
D_\mathbf{q}^0 = - U - \Lambda - 2T \cos(q_y a),
$
$
D_\mathbf{q}^x + iD_\mathbf{q}^y = - 2T \cos(q_x a) e^{iq_x a}
$
and
$
D_\mathbf{q}^z = 0,
$
where $\Lambda = 8t^2/U$ is an onsite energy offset and $T = 2t^2/U$ is the 
effective nearest-neighbor hopping parameter.
Note that $\Lambda = 2 z_{nn} t^2/U$ where $z_{nn} = 4$ is the 
nearest-neighbor coordination number. 
Thus the two-body bands can be written as
$
E_{S \mathbf{q}} = - U - \Lambda - 2T[-S\cos(q_x a) + \cos(q_y a)].
$
Apart from a constant shift, it is pleasing to see that they together correspond 
to a single cosine band $-2T[\cos(q_x a) + \cos(q_y a)]$ in the usual (non-magnetic) 
Brillouin zone of a square lattice. This is physically expected because the 
effective magnetic flux seen by the strongly-bound pair of particles is 
$
\alpha_B = \alpha_{\uparrow B} + \alpha_{\downarrow B} = 1,
$ 
and the usual Hofstadter butterfly is known to be symmetric around 
$\alpha_{\sigma B} = 1/2$, i.e., the spectrum for $\alpha_B = 1$ is 
equivalent to the non-magnetic spectrum at $\alpha_B = 0$.
In fact, starting with $\alpha_{\uparrow B} = 1/2 = - \alpha_{\downarrow B}$, 
one can easily verify that the resultant effective Hamiltonian is identical
to the one given above.
This is simply because $h_{\sigma \mathbf{k}}$ does not depend on the
sign of $\alpha_{\sigma B} = \pm 1/2$, and always exhibits time-reversal 
symmetry~\cite{umucalilar17}.

\section{Topological phase diagram}
\label{sec:tpd}

As discussed in the introduction, topological characterization of Bloch bands 
offers a new perspective on modern band theory. Similarly it may prove useful 
to construct and characterize the topological phase diagram of the two-body bands. 
As an illustration next we apply Eq.~(\ref{eqn:Nalpha}), or equivalently 
Eq.~(\ref{eqn:HNq}), to the Haldane-Hubbard model from weak to strong couplings.

\subsection{Haldane-Hubbard model}
\label{sec:hamo}

In the original Haldane model that is introduced in Sec.~\ref{sec:ham}, the energy 
gap between the upper and lower Bloch bands closes at either K or K$^\prime$ 
valley, where
$
d_\mathbf{k}^0 = 0 = d_\mathbf{k}^x = d_\mathbf{k}^y
$
and
$
d_\mathbf{k}^z = \delta \pm 3\sqrt{3} t' \sin \phi.
$
Here $\pm$ refers to $\mathbf{K}$ and $\mathbf{K'}$ points, respectively. 
It turns out while the system is a topological Chern insulator with Chern 
number $|C_s| = 1$ when
$
|\delta| < 3\sqrt{3} |t' \sin \phi|,
$ 
it is a trivial insulator with $|C_s| = 0$ when $|\delta| > 3\sqrt{3} |t' \sin \phi|$.
Thus a topological transition occurs at $|\delta| = 3\sqrt{3} |t' \sin \phi|$, 
i.e., when there is a band crossing in the system~\cite{haldane88}. 

\begin{figure}[!htb]
    \centering
    \includegraphics[width=1.\columnwidth]{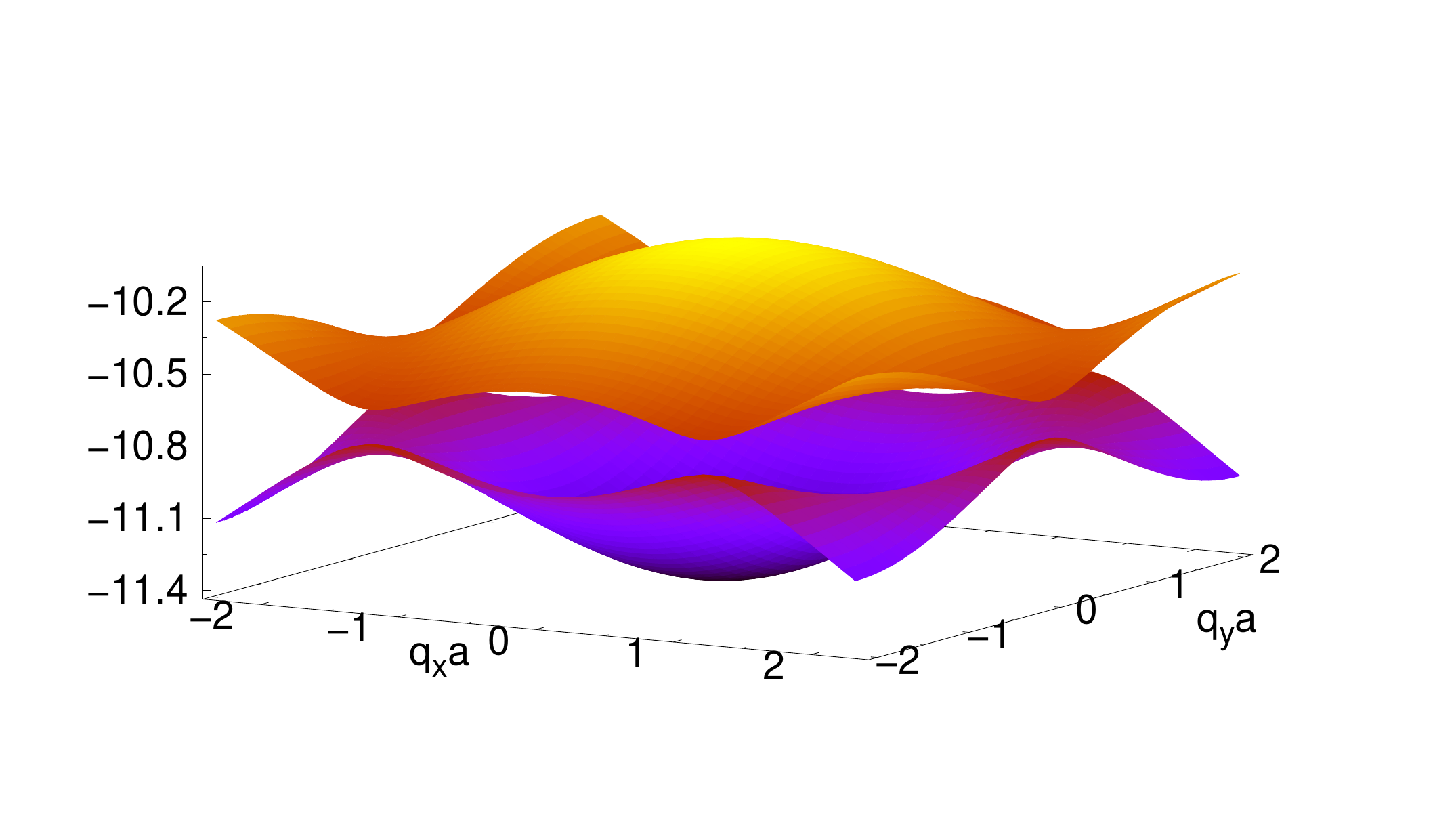}
    \caption{
    Upper and lower two-body bands correspond to $E_{+, \mathbf{q}}$ and $E_{-, \mathbf{q}}$,
    respectively, for the Haldane-Hubbard model in units of $t$. The band gap occurs 
    either at point $\mathbf{K}$ or $\mathbf{K'}$. Here $t'/t = 0.2$, $\phi = \pi/4$ 
    and $U = 10t$. 
    }
    \label{fig:2body} 
\end{figure} 

As illustrated in Fig.~\ref{fig:2body}, the two-body bands look very similar to the 
underlying Bloch bands. Accordingly the topological phase diagram of the two-body 
bands can also be traced by keeping track of their gap closings at the K and K$^\prime$ 
valleys. For instance, in the strong-coupling regime when $U/t \gg 1$, one finds 
$
D_\mathbf{q}^0 = 0 = D_\mathbf{q}^x = D_\mathbf{q}^y
$
but
$
D_\mathbf{q}^z = \Delta \pm 3\sqrt{3} T' \sin \Phi
$
at the $\mathbf{K}$ and $\mathbf{K'}$ points, respectively. 
Thus, analogous to the underlying Bloch bands, while the paired system is 
expected to be a topological Chern insulator with $|C_S| = 1$ when 
$|\Delta| < 3\sqrt{3} |T' \sin \Phi|$, it is expected to be a trivial insulator 
with $|C_S| = 0$ when $|\Delta| > 3\sqrt{3} |T' \sin \Phi|$. The topological 
transition is expected to occur at $|\Delta| = 3\sqrt{3} |T' \sin \Phi|$.

\begin{figure*}[!htb]
    \centering
    \includegraphics[width=1.8\columnwidth]{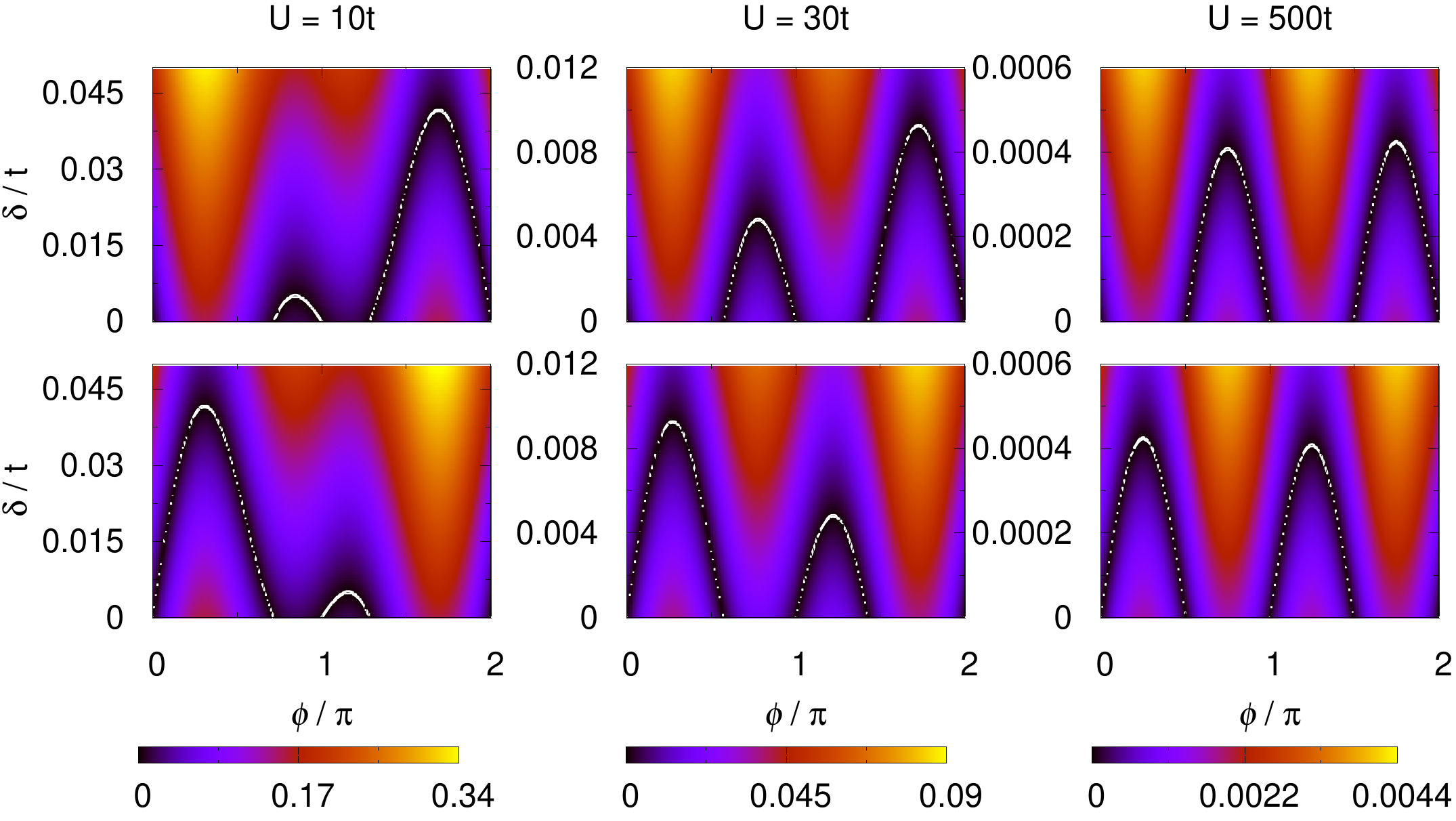}
    \caption{
    The local band gap $E_{+,\mathbf{q}} - E_{-,\mathbf{q}}$ (in units of $t$) 
    between the two-body bands as a function of phase $\phi$ (in units of $\pi$) 
    of the next-nearest-neighbor hopping $t'$ and the onsite energy difference 
    $\delta$ (in units of $t$) between sublattices. Upper and lower rows 
    correspond to the local band gaps at point $\mathbf{K}$ and $\mathbf{K'}$, 
    respectively, when $U/t = \{10,30,500\}$. Here $t'/t = 0.2$ is fixed in all 
    figures. The local band gaps vanish along the white-dotted contours within 
    the narrow black strips. For any given $U$, since the global band gap is minimum 
    of the two, increasing $\delta$ closes and reopens the band gap either at point 
    $\mathbf{K}$ or $\mathbf{K'}$ depending on $\phi$. Thus superposition of white 
    contours determines the critical boundary for the topological phase transition~\cite{salerno18}.
    When $U = 500t$ and $\delta = 0$, the Chern numbers of the upper and lower 
    two-body bands change from $\pm 1$ in $0 < \phi < \pi/2$, 
    to $\mp 1$  in $\pi/2 < \phi < \pi$,
    to $\pm 1$  in $\pi < \phi < 3\pi/2$, and
    to $\mp 1$  in $3\pi/2 < \phi < 2\pi$.
    } 
    \label{fig:haldane} 
\end{figure*} 

In Fig.~\ref{fig:haldane} we present the local band gap $E_{+,\mathbf{q}} - E_{-,\mathbf{q}}$
between the two-body bands as a function of phase $\phi$ of the 
next-nearest-neighbor hopping $t'$ and the onsite energy difference $\delta$ between 
sublattices. Here we set $t'/t = 0.2$ in all figures, where the upper and lower rows 
correspond to the local band gaps at points $\mathbf{q} =  \mathbf{K}$ and $\mathbf{q} = \mathbf{K'}$, 
respectively, and different columns correspond to $U/t = \{10,30,500\}$. 
The local band gaps vanish along the white-dotted contours within the narrow black strips. 
Since the global band gap is minimum of the local band gaps, superposition of white 
contours determines the location of the vanishing global band gap.
For a given $\mathbf{\mathbf{K}}$ or $\mathbf{K'}$, each strip has one primary (i.e., larger) 
and one secondary (i.e., smaller) lobe as a function of $0 \le \phi < 2\pi$. 
The secondary lobes are as large as the primary ones only in the strong-coupling regime, 
and this is in perfect agreement with our strong-coupling analysis. 

In the case of $U = 500t$, the period $\pi/2$ and the amplitude 
$
3\sqrt{3}(0.2)^2t/500 \approx 0.00042t
$ 
of the oscillation that is produced by the combined strips (i.e., superposition of the 
white-dotted contours that are shown in $\mathbf{K}$ and $\mathbf{K'}$) match very well 
with our gap closing condition 
$
\delta = 3\sqrt{3} |\sin (2\phi)|t'^2/U
$ 
derived above. However, in the case of $U = 30t$, the oscillation of the 
combined strips has a period of $2\pi$, and the amplitudes of the primary and 
secondary lobes deviate substantially from our strong-coupling prediction $0.0069t$. 
This deviation shows that the higher-order corrections to the effective Hamiltonian
play a crucial role in determining the phase boundary.  
It is pleasing to see that our exact results coming out of Eq.~(\ref{eqn:Nalpha}) 
are in full agreement with the recent results that are based on exact diagonalization 
in real space~\cite{salerno18}. There the effective Hamiltonian is derived up to 
third order in $1/U$, showing that the two-body bound states are described by a 
generalized Haldane model in general~\footnote{
We are not aware of any study on the topological phase diagram of the generalized 
Haldane model, where next-nearest-neighbor (i.e., intra-sublattice) hoppings are 
different for different sublattices, to compare our two-body results.}. 
When $U$ is finite, the amplitudes and phases of the effective next-nearest-neighbor 
hopping parameter turn out to be different for different sublattices, 
i.e., one has to consider
$
T_{nnn} = T'_{AA} e^{i \Phi_A}
$ 
and 
$
T_{nnn} = T'_{BB} e^{i \Phi_B}
$
for intra-sublattice hoppings where $T'_{AA} \ne T'_{BB}$ and $\Phi_A \ne \Phi_B$
have much more complicated dependences on $t, t', \phi, \delta$ and $U$~\cite{salerno18}.

Furthermore Fig.~\ref{fig:haldane} shows that the secondary lobes already become 
small when the interaction is lowered down to $U = 10t$. As they start disappearing 
towards the weakly-interacting regime (e.g., when $U \lesssim 5t$ not shown), 
the lobe structure resembles to that of Bloch bands, i.e., the primary lobe of 
$\mathbf{K}$ extents from $\phi = 0$ to $\pi$ and that of $\mathbf{K'}$ extents from 
$\pi$ to $2\pi$. This is in such a way that the oscillation of the combined strips 
has a period of $\pi$. Thus both the amplitude and the width of the primary lobes 
grow in size with decreasing interactions. Next we calculate the Chern number 
of the two-body bands in the strong-coupling regime and show that it is consistent 
with the resultant lobe structure.

\subsection{Chern number}
\label{sec:cn}

The Berry curvature $\Omega_{S \mathbf{q}}$ of the two-body eigenvector $\mathbf{F}_{S \mathbf{q}}$ 
is derived in the Appendix B. Under the restrictive assumption that the two-body bound 
states $|S \mathbf{q} \rangle$ are orthonormal to each other, i.e., 
when the identity operator
$
\mathcal{I} = |+, \mathbf{q} \rangle \langle +, \mathbf{q}| 
+ |-, \mathbf{q} \rangle \langle -, \mathbf{q}|
$
is approximately satisfied, for every $\mathbf{q}$, there we show that
\begin{align}
\Omega_{S \mathbf{q}} = -2\mathrm{Im} \frac{
\langle S \mathbf{q}| \partial_x H_{S \mathbf{q}} |-S, \mathbf{q} \rangle
\langle -S, \mathbf{q}| \partial_y H_{S \mathbf{q}} |-S, \mathbf{q} \rangle
}
{(E_{S \mathbf{q}} - E_{-S, \mathbf{q}})^2}.
\end{align}
Here $H_{S \mathbf{q}}$ is defined by Eq.~(\ref{eqn:HSq}) and $\partial_j$ stands 
for $\partial/\partial q_j$. 
Then the Chern number of the two-body bands is given by the usual expression
$
C_S = \frac{2\pi}{L^2} \sum_\mathbf{q} \Omega_{S \mathbf{q}},
$
where $L_x = L_y = L \gg a$ is the side-length of the square-shaped lattice. 
In the case of Haldane model, $L$ and $M_c$ are such that $M_c = 2L^2/(3\sqrt{3}a^2)$, 
i.e., dividing the area $8\pi^2/(3\sqrt{3}a^2)$ of the hexagon-shaped Brillouin 
zone to the area $4\pi^2/L^2$ per $\mathbf{q}$ state gives the number $M_c$ 
of $\mathbf{q}$ states (per band) in the Brillouin zone. 
Unlike that of the Bloch bands, we note that $\Omega_{+, \mathbf{q}}$ and 
$- \Omega_{-, \mathbf{q}}$ are not necessarily equal to each other by construction
because $H_{+, \mathbf{q}}$ and $H_{-, \mathbf{q}}$ are different.

Our numerical calculations show that the orthonormality condition is well-satisfied 
in the strong-coupling regime. 
For instance when $\delta = 0$ and $\phi = \pi/4$, we find that the inner product
$
|\langle +, \mathbf{q}| -, \mathbf{q} \rangle|
$
is bounded approximately by 
$\{3 \times 10^{-8}, 10^{-4}, 7 \times 10^{-4}, 2 \times 10^{-3}\}$ 
(i.e., for every $\mathbf{q}$ in the Brillouin zone) when 
$U/t = \{500, 30, 15, 10\}$, respectively. 
The corresponding Chern numbers for the upper ($S = +$) and 
lower ($S = -$) two-body bands are
$C_+ = 0.994$ and $C_- = -0.994$ for $U = 500t$,
$C_+ = 1.03$ and $C_- = -1.02$ for $U = 30t$,
$C_+ = 1.15$ and $C_- = -1.09$ for $U = 15t$, and
$C_+ = 1.42$ and $C_- = -1.21$ for $U = 10t$.
When $\phi = 7\pi/4$, we confirm that all of these Chern numbers simply 
change signs with exactly the same magnitudes.
Similarly when $\delta = 0$ and $\phi = 3\pi/4$, 
we find that the inner product is bounded approximately by 
$\{3 \times 10^{-8}, 10^{-4}, 10^{-3}, 3 \times 10^{-3}\}$ 
when $U/t = \{500, 30, 15, 10\}$, respectively.
The corresponding Chern numbers are
$C_+ = -0.994$ and $C_- = 0.994$ for $U = 500t$,
$C_+ = -1.02$ and $C_- = 1.01$ for $U = 30t$,
$C_+ = -1.08$ and $C_- = 1.05$ for $U = 15t$, and
$C_+ = -0.75$ and $C_- = 0.72$ for $U = 10t$.
When $\phi = 5\pi/4$, we again confirm that these Chern numbers also 
change signs with exactly the same magnitudes.

As long as lowering $U/t$ from 500 to 10 does not open or close any energy gap, 
e.g., Fig.~\ref{fig:haldane} shows that this is the case when $\delta = 0$ and 
$\phi = \{\pi/4, 3\pi/4, 5\pi/4, 7\pi/4\}$, the associated Chern number $C_S$ 
cannot change and must remain invariant for a given lobe. This is not the case
in our numerical calculations because the orthonormality condition progressively 
fails more and more at lower $U$ values. Thus our approach is by construction 
not expected to reproduce the correct $|C_S| = 1$ in the weak-coupling regime. 
On the other hand, we use roughly $2000$ mesh points and distribute them uniformly 
in the Brillouin zone in our numerical calculations, and increasing the mesh 
size may give slightly better results in the strong-coupling regime, e.g., when 
$U= 500t$ or $U = 30t$. This is because since the Berry curvature $\Omega_{S \mathbf{q}}$ 
makes a much larger contribution to $C_S$ in the vicinity of $\mathbf{K}$ and 
$\mathbf{K'}$ points, increasing the mesh size must eventually give $|C_S| = 1$ 
up to a very high precision once the effective Hamiltonian discussed in 
Sec.~\ref{sec:ham} becomes applicable.

\section{Conclusion}
\label{sec:conc}

In summary here we studied the two-body problem in a Haldane-Hubbard model, 
and constructed its topological phase diagrams as a function of interaction
strength, by keeping track of the gap closings in between the two-body bands. 
For a given center of mass momentum,
the two-body bands are determined by a nonlinear eigenvalue problem, and 
its self-consistent solutions are obtained numerically through an iterative 
approach. We found that while the lobe structure of the weakly-interacting 
phase diagram resembles to that of the Bloch bands, two additional lobes 
appear and grow gradually with increasing interactions. Our strong-coupling 
analysis is in perfect agreement with the topological phase diagram, 
where an effective Hamiltonian is derived for the two-body bands through 
perturbation theory. In addition, assuming that the eigenvectors of the 
nonlinear eigenvalue problem form an orthonormal set, we reformulated the 
Berry cuvature and the associated Chern number. 
This assumption is typically fulfilled in the strongly-interacting regime, 
where, e.g., the resultant Chern numbers are again consistent with the lobe 
structure in the Haldane-Hubbard model. 
As an outlook, calculation of the correct Chern numbers in the weak-coupling 
regime is indeed an interesting area of investigation. This may be achieved
via an alternative formulation that is based only on one of eigenvectors 
of the nonlinear eigenvalue problem without an explicit reference to the 
other eigenvectors or to the orthonormality condition. For instance 
Ref.~\cite{fukui05} offers such a promising approach.

\begin{acknowledgments}
The author acknowledges funding from T{\"U}B{\.I}TAK.
\end{acknowledgments}

\appendix
\subsection*{Appendix A: Expansion coefficients in Sec.~\ref{sec:sbr}}

In the strong-coupling regime when the binding energy is much larger than the 
single-particle energies, i.e., when $|x| \ll 1$ with
$
x = (\varepsilon_{s \uparrow \mathbf{k}+\mathbf{q}} 
+ \varepsilon_{s' \downarrow -\mathbf{k}})/E_{S \mathbf{q}},
$
the matrix elements given in Eq.~(\ref{eqn:Hmatrix}) can be expanded as a geometric 
series using
$
1/(1 - x) = 1  + x + x^2 + \cdots.
$
For instance, in the expansion of the diagonal elements $H_{S\mathbf{q}}^{AA}$ 
and $H_{S\mathbf{q}}^{BB}$, the zeroth-order terms follow from
$
\sum_{ss'} |s_{A \uparrow \mathbf{k+q}}|^2 |s'_{A \downarrow -\mathbf{k}}|^2  = 
1 = \sum_{ss'} |s_{B \uparrow \mathbf{k+q}}|^2 |s'_{B \downarrow -\mathbf{k}}|^2,
$
the first-order terms follow from
$
\sum_{ss'} |s_{A \uparrow \mathbf{k+q}}|^2 |s'_{A \downarrow -\mathbf{k}}|^2
(\varepsilon_{s \uparrow \mathbf{k}+\mathbf{q}} + \varepsilon_{s' \downarrow -\mathbf{k}}) 
= d_{\uparrow \mathbf{k+q}}^0 + d_{\uparrow \mathbf{k+q}}^z 
+ d_{\downarrow -\mathbf{k}}^0 + d_{\downarrow -\mathbf{k}}^z
$
and
$
\sum_{ss'} |s_{B \uparrow \mathbf{k+q}}|^2 |s'_{B \downarrow -\mathbf{k}}|^2
(\varepsilon_{s \uparrow \mathbf{k}+\mathbf{q}} + \varepsilon_{s' \downarrow -\mathbf{k}}) 
= d_{\uparrow \mathbf{k+q}}^0 - d_{\uparrow \mathbf{k+q}}^z 
+ d_{\downarrow -\mathbf{k}}^0 - d_{\downarrow -\mathbf{k}}^z,
$
and the second-order terms follow from
$
\sum_{ss'} |s_{A \uparrow \mathbf{k+q}}|^2 |s'_{A \downarrow -\mathbf{k}}|^2
(\varepsilon_{s \uparrow \mathbf{k}+\mathbf{q}} + \varepsilon_{s' \downarrow -\mathbf{k}})^2
= (d_{\uparrow \mathbf{k+q}}^0 + d_{\downarrow -\mathbf{k}}^0) 
+ d_{\uparrow \mathbf{k+q}}^2 + d_{\downarrow -\mathbf{k}}^2
+ 2d_{\uparrow \mathbf{k+q}}^0 (d_{\uparrow \mathbf{k+q}}^z + d_{\downarrow -\mathbf{k}}^z)
+ 2d_{\downarrow -\mathbf{k}}^0 (d_{\uparrow \mathbf{k+q}}^z + d_{\downarrow -\mathbf{k}}^z)
+ 2d_{\uparrow \mathbf{k+q}}^z d_{\downarrow -\mathbf{k}}^z
$
and
$
\sum_{ss'} |s_{B \uparrow \mathbf{k+q}}|^2 |s'_{B \downarrow -\mathbf{k}}|^2
(\varepsilon_{s \uparrow \mathbf{k}+\mathbf{q}} + \varepsilon_{s' \downarrow -\mathbf{k}})^2
= (d_{\uparrow \mathbf{k+q}}^0 + d_{\downarrow -\mathbf{k}}^0) 
+ d_{\uparrow \mathbf{k+q}}^2 + d_{\downarrow -\mathbf{k}}^2
- 2d_{\uparrow \mathbf{k+q}}^0 (d_{\uparrow \mathbf{k+q}}^z + d_{\downarrow -\mathbf{k}}^z)
- 2d_{\downarrow -\mathbf{k}}^0 (d_{\uparrow \mathbf{k+q}}^z + d_{\downarrow -\mathbf{k}}^z)
+ 2d_{\uparrow \mathbf{k+q}}^z d_{\downarrow -\mathbf{k}}^z.
$
Here we made frequent use of the normalization condition
$
u_{\sigma \mathbf{k}}^2 + v_{\sigma \mathbf{k}}^2 = 1
$ 
for the Bloch eigenvectors. Similarly, in the expansion of the off-diagonal elements 
$H_{S \mathbf{q}}^{AB} = (H_{S\mathbf{q}}^{BA})^*$, the trivial zeroth-order terms follow from
$
\sum_{ss'} s^*_{A \uparrow \mathbf{k+q}} s'^*_{A \downarrow -\mathbf{k}}
s_{B \uparrow \mathbf{k+q}} s'_{B \downarrow -\mathbf{k}} = 0,
$
the trivial first-order terms follow from
$
\sum_{ss'} s^*_{A \uparrow \mathbf{k+q}} s'^*_{A \downarrow -\mathbf{k}}
s_{B \uparrow \mathbf{k+q}} s'_{B \downarrow -\mathbf{k}}
(\varepsilon_{s \uparrow \mathbf{k}+\mathbf{q}} + \varepsilon_{s' \downarrow -\mathbf{k}}) 
= 0,
$
and the non-trivial second-order terms follow from
$
\sum_{ss'} s^*_{A \uparrow \mathbf{k+q}} s'^*_{A \downarrow -\mathbf{k}}
s_{B \uparrow \mathbf{k+q}} s'_{B \downarrow -\mathbf{k}}
(\varepsilon_{s \uparrow \mathbf{k}+\mathbf{q}} + \varepsilon_{s' \downarrow -\mathbf{k}})^2
= 2 g_{\uparrow \mathbf{k+q}}  g_{\downarrow -\mathbf{k}}.
$
Here we note that
$
s^*_{A \uparrow \mathbf{k+q}} s'^*_{A \downarrow -\mathbf{k}}
s_{B \uparrow \mathbf{k+q}} s'_{B \downarrow -\mathbf{k}} = 
- g_{\uparrow \mathbf{k+q}}  g_{\downarrow -\mathbf{k}} (-1)^{\delta_{ss'}} 
/ (4 d_{\uparrow \mathbf{k+q}}  d_{\downarrow -\mathbf{k}})
$
in such a way that 
$
\sum_{ss'} (-1)^{\delta_{ss'}} = 0
$
leads to the zeroth-order result,
$
\sum_{ss'} (\varepsilon_{s \uparrow \mathbf{k}+\mathbf{q}} 
+ \varepsilon_{s' \downarrow -\mathbf{k}})(-1)^{\delta_{ss'}} = 0
$
leads to the first-order result, and
$
\sum_{ss'} (\varepsilon_{s \uparrow \mathbf{k}+\mathbf{q}} + \varepsilon_{s' \downarrow -\mathbf{k}})^2
(-1)^{\delta_{ss'}} = - 8 d_{\uparrow \mathbf{k+q}}  d_{\downarrow -\mathbf{k}}
$
leads to the second-order result.

\subsection*{Appendix B: Berry curvature in Sec.~\ref{sec:cn}}

Unlike the one-body problem, the two-body bands $E_{N \mathbf{q}}$ and their 
corresponding eigenstates $| N \mathbf{q} \rangle$ are determined by the nonlinear 
eigenvalue problem given in Eq.~(\ref{eqn:HNq}). For this reason the standard 
formulation of the Berry curvature is not applicable here.
In this Appendix we formulate the Berry curvature of the two-body bands by 
following closely the footsteps of Berry in his seminal paper~\cite{berry84}. 
However, it is important to emphasize that the derivation below is not general, 
and it only applies when the self-consistent solutions for the 
eigenvectors $|N \mathbf{q} \rangle$ form an orthonormal set, 
i.e., when the identity operator
$
\mathcal{I} = \sum_N |N \mathbf{q} \rangle \langle N \mathbf{q}| 
$
is satisfied, for every $\mathbf{q}$. Our numerical calculations suggest that this 
condition can approximately be satisfied in the strong-coupling 
(i.e., $|U|/t \gg 1$) regime once the bound states localize strongly on one of the 
sublattices. In addition our numerical calculations show that it is also approximately 
satisfied in the weak-coupling (i.e., $|U|/t \to 0$) limit when the lowest or highest 
Bloch band is flat. See also Appendix C below. 

The Berry connection of the two-body states is defined as
$
\mathbf{\mathcal{A}}_{N\mathbf{q}} = i \langle N \mathbf{q} | \boldsymbol{\nabla} N \mathbf{q} \rangle
= - \mathrm{Im} \langle N \mathbf{q} | \boldsymbol{\nabla} N \mathbf{q} \rangle,
$
where the second equality follows because the inner product is an imaginary number 
due to 
$
\boldsymbol{\nabla}  \langle N \mathbf{q} | N \mathbf{q} \rangle = 0
$
for the normalized states. Then the Berry curvature of the two-body states is
$
\boldsymbol{\Omega}_{N \mathbf{q}} = \boldsymbol{\nabla} \times \mathbf{\mathcal{A}}_{N\mathbf{q}}
 = - \mathrm{Im} \langle \boldsymbol{\nabla} N \mathbf{q} | \times | \boldsymbol{\nabla} N \mathbf{q} \rangle,
$
where the cross product is between three components of bra and ket vectors.
By acting $\boldsymbol{\nabla}$ on the nonlinear eigenvalue Eq.~(\ref{eqn:HNq}), 
one obtains
$
(H_{N \mathbf{q}} - E_{N \mathbf{q}}) | \boldsymbol{\nabla} N \mathbf{q} \rangle
= (\boldsymbol{\nabla} E_{N \mathbf{q}} - \boldsymbol{\nabla}H_{N \mathbf{q}}) |N \mathbf{q} \rangle.
$
Assuming no band crossings, and given that 
$
\langle N \mathbf{q} | \boldsymbol{\nabla} H_{N \mathbf{q}} |N \mathbf{q} \rangle
= \boldsymbol{\nabla} E_{N \mathbf{q}},
$
the right-hand side of the previous expression does not have any projection onto 
$|N \mathbf{q} \rangle$. Thus one can safely act on it with 
$(H_{N \mathbf{q}} - E_{N \mathbf{q}})^{-1}$, and determine 
$
| \boldsymbol{\nabla} N \mathbf{q} \rangle.
$
This leads to 
$
\boldsymbol{\Omega}_{N \mathbf{q}} = -\mathrm{Im} 
\langle N \mathbf{q} | 
(\boldsymbol{\nabla} E_{N \mathbf{q}}-\boldsymbol{\nabla} H_{N \mathbf{q}}) 
(H_{N \mathbf{q}} - E_{N \mathbf{q}})^{-1} \times (H_{N \mathbf{q}} - E_{N \mathbf{q}})^{-1}
(\boldsymbol{\nabla} E_{N \mathbf{q}}-\boldsymbol{\nabla} H_{N \mathbf{q}}) 
|N \mathbf{q} \rangle.
$
Then, by plugging the identity operator $\mathcal{I}$ across the cross product, 
and noting that the left and right sides of the cross product do not have any 
projections onto $\langle N \mathbf{q}|$ and $|N \mathbf{q} \rangle$, respectively, 
one finds
\begin{align}
\boldsymbol{\Omega}_{N \mathbf{q}} = -\mathrm{Im} \sum_{M \ne N}
\frac{
\langle N \mathbf{q}| \boldsymbol{\nabla} H_{N \mathbf{q}} |M \mathbf{q} \rangle
\times
\langle M \mathbf{q}| \boldsymbol{\nabla} H_{N \mathbf{q}} |N \mathbf{q} \rangle
}
{(E_{M \mathbf{q}} - E_{N \mathbf{q}})^2},
\end{align}
through some simple algebra. In particular, for a two-dimensional system lying in 
the $xy$ plane, e.g., in the Haldane-Hubbard model, 
$
\boldsymbol{\Omega}_{N \mathbf{q}} = \Omega_{N \mathbf{q}} \boldsymbol{\widehat{k}}
$
is along the $z$ direction.

\subsection*{Appendix C: Isolated flat bands}

Here we consider a number of weakly-coupled degenerate dispersionless flat bands that 
are energetically isolated from the rest of the Bloch bands in the spectrum~\cite{herzog22}. 
Suppose 
$
\varepsilon_{n \sigma \mathbf{k}} = 0
$ 
is the energy of these flat bands, and they are separated by an energy $\varepsilon_0$ 
from the nearest band. In this case Eq.~(\ref{eqn:Hmatrix}) reduces to
$
H^{\alpha \beta}_{N \mathbf{q}} \to H^{\alpha \beta}_{\mathbf{q}} 
$
where
\begin{align}
H^{\alpha \beta}_{\mathbf{q}} = -\frac{U} {M_c} \sum_{nm\mathbf{k}}
n_{\beta \uparrow \mathbf{k+q}}^* m_{\beta \downarrow -\mathbf{k}}^*
m_{\alpha \downarrow -\mathbf{k}} n_{\alpha \uparrow \mathbf{k+q}}.
\end{align}
This effective Hamiltonian is valid only in the $U/\varepsilon_0 \to 0$ limit so that 
the dispersive bands can be projected out of the system. In the presence of time-reversal 
symmetry, i.e., when 
$
n_{\alpha \downarrow -\mathbf{k}}^* = n_{\alpha \uparrow \mathbf{k}} \equiv n_{\alpha \mathbf{k}} 
$
and 
$
\varepsilon_{n \downarrow -\mathbf{k}} = \varepsilon_{n \uparrow \mathbf{k}} \equiv \varepsilon_{n \mathbf{k}},
$
we note that $H^{\alpha \beta}_{\mathbf{q}}$ is precisely the exact many-body 
Hamiltonian Eq.~(13) that is derived in Ref.~\cite{herzog22} under the same settings.
This coincidence suggests that the interaction between the resultant two-body bound 
states, i.e., Cooper pairs, is negligible. Furthermore, since $H^{\alpha \beta}_{\mathbf{q}}$ 
does not depend on $E_{N \mathbf{q}}$ in this particular setting, the orthonormality condition 
is automatically intact for the resultant eigenvectors.

\bibliography{refs}

\end{document}